\documentclass[aps,prb,amsmath,amssymb,a4paper,reprint,nofootinbib,showkeys]{revtex4-2}

\usepackage{verbatim}
\usepackage{graphicx}
\usepackage[utf8]{inputenc}
\usepackage{hyperref}
\usepackage{epstopdf}
\usepackage{listings}
\usepackage{xcolor}
\definecolor{blue}{rgb}{0,0,1}
\newcommand*\blue{\color{blue}}
\usepackage{array}

\newcommand{\be}{\begin{equation}}
\newcommand{\ee}{\end{equation}}
\newcommand{\md}{\mathrm{d}}
\newcommand{\nn}{\nonumber}
\newcommand{\eF}{\epsilon_\mathrm{_F}}
\newcommand{\Fref}[1]{Fig.~\ref{#1}}
\newcommand{\Eqref}[1]{Eq.~(\ref{#1})}

\begin{document}

\title[Zero sound with ferromagnetic exchange interaction]
{Possible zero sound in layered perovskites with ferromagnetic 
\textit{s-d} exchange interaction}

\author{Todor~M.~Mishonov}
\email{mishonov@bgphysics.eu}
\affiliation{Georgi Nadjakov Institute of Solid State Physics, Bulgarian
Academy of Sciences, 72 Tzarigradsko Chauss\'ee, BG-1784 Sofia, Bulgaria}

\author{Nedelcho~I.~Zahariev}
\email{zahariev@issp.bas.bg}
\affiliation{Georgi Nadjakov Institute of Solid State Physics, Bulgarian
Academy of Sciences, 72 Tzarigradsko Chauss\'ee, BG-1784 Sofia, Bulgaria}

\author{Hassan Chamati}
\email{chamati@issp.bas.bg}
\affiliation{Georgi Nadjakov Institute of Solid State Physics, Bulgarian
Academy of Sciences, 72 Tzarigradsko Chauss\'ee, BG-1784 Sofia, Bulgaria}

\author{Albert~M.~Varonov}
\email{varonov@issp.bas.bg}
\affiliation{Georgi Nadjakov Institute of Solid State Physics, Bulgarian
Academy of Sciences, 72 Tzarigradsko Chauss\'ee, BG-1784 Sofia, Bulgaria}

\date{1 August 2022, 17:47}

\begin{abstract}
We analyze the conditions for observation of zero sound in layered perovskites 
with transition metal ion on chalcogenide oxidizer. 
We conclude that propagation of zero sound is possible only for
a ferromagnetic sign of the \textit{s-d} interaction.
If the \textit{s-d} exchange integral \textit{J}$\!_{sd}$ has antiferromagnetic sign,
as it is perhaps in the case for layered cuprates,
zero sound is a thermally activated dissipation
mode, which generates only ``hot spots'' in the Angle Resolved Photoemission Spectroscopy (ARPES) data along the Fermi contour.
We predict that zero sound will be observable for transition metal
perovskites with 4\textit{s} and 3\textit{d} levels close to the
\textit{p}-level of the chalcogenide. The simultaneous
lack of superconductivity, the appearance of hot spots in ARPES data, and the
proximity of the three named levels, represents the significant hint for the
choice of material to be investigated.
\end{abstract}


\maketitle

\section{Introduction}

The theoretical prediction of zero sound by Landau~\cite{Landau_theory,Landau_zero_sound}
and subsequent experimental observations~\cite{Keen:65,Abel:66} in $^3$He 
was a powerful evidence
of the applicability of Landau picture of Fermi quasi-particles excitations 
and their self-consistent motion to (strongly correlated) Fermi liquids.\footnote{This work has to be cited  as Ref.~\cite{sound2022}}
The zero Fermi sound in metals, more precisely zero spin sound, was observed in 
Cr metal~\cite{Fukuda:96,Endoh:97}.
These experiments stimulated studies in the framework of the Hubbard
model. Fuseya \textit{et al.}~\cite{Fuseya:00} reached the important, to our
further analysis, conclusion that Landau parameter, \textit{i.e.} the averaged on the Fermi surface 
Fermi liquid interaction kernel $f(\mathbf{p},\mathbf{q})$, can change its sign close to half filling of the conduction band.
Later on Tsuruta~\cite{Tsuruta:10} used two dimensional $t$-$t^\prime$ Hubbard
model to study zero spin sound in antiferromagnetic metals.
Here, we consider that this approach can be useful 
to study zero sound propagation and its importance from a materials science perspective.
The purpose of the present paper is to explore the possibility
of zero sound propagation in a layered perovskite with the structure of
high-$T_c$ having a CuO$_2$ conduction plane.

Postulating the interaction kernel
$f(\mathbf{p},\mathbf{q})$~\cite[Eq.~(2.1)]{LL9}
between electrons with different momenta $\mathbf{p}$ and $\mathbf{q}$,
we can explain different electronic phenomena in superconductivity and magnetism.
The Fermi liquid approach provides results for the 
magnetic susceptibility, heat capacity, and effective masses.
For illustration, in many cases  the interaction  is modeled 
by a separable kernel 
$f(\mathbf{p},\mathbf{q})\propto\chi(\mathbf{p})\chi(\mathbf{q})$
and for our Hamiltonian separability holds.

The $s$-$d$ exchange lies in the origin of the magnetic properties of
transition metal compounds.
Its most usual version was proposed by Schubin and
Wonsovsky~\cite{Shubin}, Zener~\cite{Zener1,Zener2,Zener3} and 
Kondo~\cite{Kondo}.
%
The purpose of the present study is to analyze whether the $s$-$d$ exchange can lead
to propagation of zero sound in transition metal compounds.

We anticipate here, that an anti-ferromagnetic sign of
the $s$-$d$ coupling $J_{sd}$
leads to a singlet superconductivity while
a ferromagnetic sign of $J_{sd}$ is able to explain the repulsion necessary for the propagation 
of zero sound.
We wish to emphasize that zero  sound has not yet been observed 
in normal metals~\cite{Abrikosov}.
This may be traced back to the fact
that the $s$-$d$ exchange interaction was not used as a guide for 
the choice of appropriate materials.

In the next section we introduce the notions and notations developed to explain the 
electronic properties of the CuO$_2$ plane and its superconductivity, and analyze
how a similar Hamiltonian would be used to predict zero sound in layered cuprates.
Finally, we conclude that layered transition metal compounds may serve
as the best candidates to search for zero sound in normal metals.

\section{The \textit{s-d} LCAO Hamiltonian for the CuO$_2$ plane in
momentum $\mathbf{p}$-representation}
\begin{figure}[t!]
\centering
\includegraphics[width=\columnwidth]{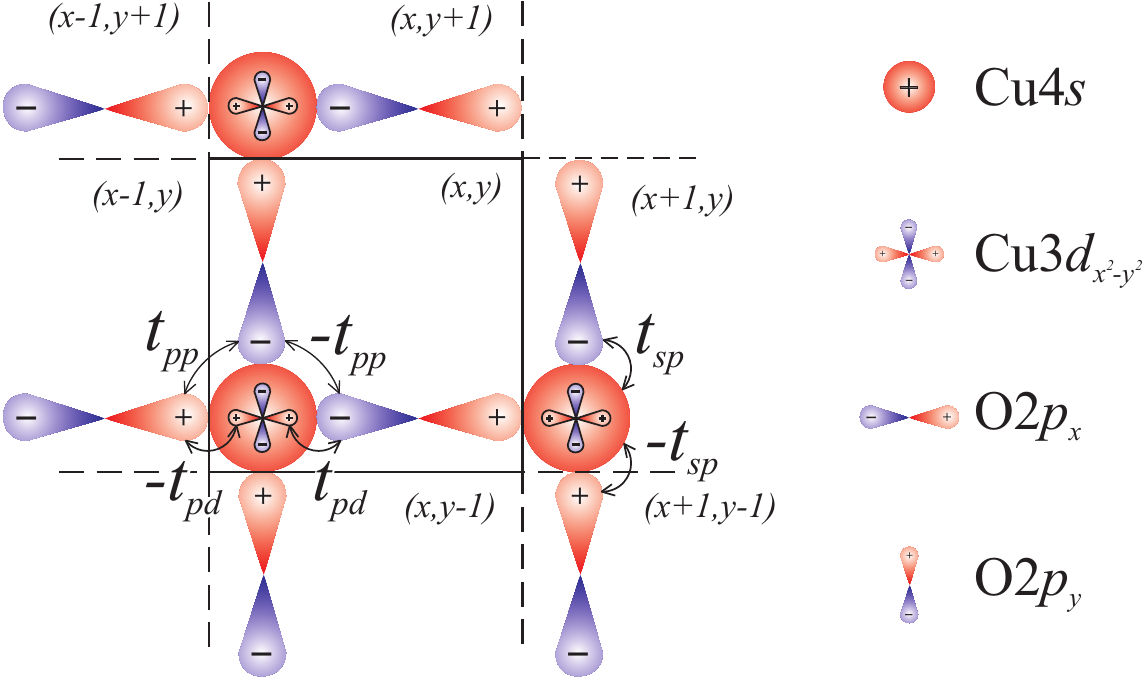}
\caption{
Single electron processes in the conducting CuO$_2$ plane of layered cuprates after Ref.~\cite[Fig.~1.1]{MishonovPenev:11}.
LCAO is the basis of the chemical intuition.
We have Hilbert space spanned over Cu$4s$, Cu$3d_{x^2-y^2}$, O$2p_x$, and 
O$2p_y$ states. 
The LCAO Hamiltonian is parameterized by transfer integrals 
$t_{pd}$,
$t_{sp}$, and $t_{pp}$ and single site energies 
$\epsilon_d$, $\epsilon_s$, $\epsilon_p$.}
\label{Fig:LCAO}
\end{figure}

The Hamiltonian in the $\mathbf{r}$-representation is given in
Ref.~\cite[Eq.~(1.2)]{MishonovPenev:11}, here we start with the
Hamiltonian in the $\mathbf{p}$-representation
\begin{align}
\hat H^\prime=&\sum_{\mathbf{p}, \, \alpha}
\hat{\Psi}_\mathbf{p}^\dagger 
\left(H_\mathrm{LCAO}-\mu\mathbb{1} \right)
\hat{\Psi}_\mathbf{p}
\nn\\
& -  \frac{J_{sd}}{\mathcal N} \sum_{\substack{\mathbf{p}^\prime+\mathbf{q}^\prime
=\mathbf{p}+\mathbf{q} \\ \alpha, \, \beta}} 
\hat{S}_{\mathbf{q}^\prime,\beta}^\dagger\hat{D}_{\mathbf{p}^\prime,\alpha}^\dagger
\hat{S}_{\mathbf{p},\alpha}\hat{D}_{\mathbf{q},\beta},
\end{align}
where
\begin{equation*}
H_\mathrm{LCAO}\equiv
\begin{pmatrix}
\epsilon_d  & 0                    & t_{pd}s_x      &  -t_{pd}s_y  \\
0                & \epsilon_s    & t_{sp}s_x       & t_{sp}s_y     \\
t_{pd}s_x  & t_{sp}s_x       & \epsilon_\mathrm{p} & -t_{pp}s_xs_y \\
-t_{pd}s_y &t_{sp}s_y        & -t_{pp}s_xs_y& \epsilon_\mathrm{p}
&
\end{pmatrix},
\end{equation*}
\begin{equation*}
s_x\equiv 2\sin(p_x/2),\qquad
s_y\equiv 2\sin(p_y/2), \nn\\
\end{equation*}
\begin{equation*}
\hat{\Psi}_\mathbf{p}^\dagger\equiv
\begin{pmatrix}
\hat{D}^\dagger_{\mathbf{p},\alpha} & \hat{S}^\dagger_{\mathbf{p},\alpha} &
\hat{X}^\dagger_{\mathbf{p},\alpha} & \hat{Y}^\dagger_{\mathbf{p},\alpha}
\end{pmatrix}, \nn
\end{equation*}
and the summation is actually an integration in the momentum space
and $\mathcal{N}\gg1$ is the total number of elementary cells for which we
assume periodic boundary conditions
\begin{align}
\frac1{\mathcal N} \sum_\mathbf{p} \dots\;\;\equiv 
\iint\limits _{\{p_x, p_y\} \in (0,2\pi)} \frac{\md p_x \md p_y}{(2 \pi)^2}\dots\nn
\end{align}
where the momentum variables are dimensionless phases 
$p_x, p_y, q_x, q_y \in (0,2\pi)$;
the dimensional momentum is $\mathbf{P}=(\hbar/a_0)\,\mathbf{p}.$
In this LCAO Hamiltonian
$\epsilon_s$, $\epsilon_d$ and $\epsilon_p$ are the single site energies
of an electron in Cu4$s$, Cu3$d_{x^2-y^2}$ and O2$\mathrm p$ states,
$t_{sp}$, $t_{pd}$ and $t_{pd}$ are hopping amplitudes between
neighboring orbitals.
The $s$-$d$ interaction is parameterized by the exchange integral 
$J_{sd}$ which we consider as a perturbation.
Schematically, the atomic wave functions are depicted in \Fref{Fig:LCAO}.
The chemical potential is denoted by $\mu$ and for the operator of the 
number of electrons we have the standard expression
\begin{align}
\hat N & =-\partial_\mu \hat{H}^\prime\notag \\
& = \sum_{\mathbf{p}, \, \alpha} (
\hat D_{\mathbf{p},\alpha}^\dagger \hat D_{\mathbf{p},\alpha}
+\hat S_{\mathbf{p},\alpha}^\dagger \hat S_{\mathbf{p},\alpha} 
+\hat X_{\mathbf{p},\alpha}^\dagger \hat X_{\mathbf{p},\alpha}
+\hat Y_{\mathbf{p},\alpha}^\dagger \hat Y_{\mathbf{p},\alpha}), \nn
\end{align}
which we treat self-consistently.

\subsection{Conduction band reduction}
In order to derive the effective Hamiltonian describing the zero
sound, we perform successive conduction band reductions.
The first one is the number of particles.
For notions and notations we  
follow the description of the electronic properties of the CuO$_2$ plane.
In the hole doped phase of the CuO$_2$ plane we have
2 completely filled oxygen bands O$2p$
and one partially filled Cu3$d_{x^2-y^2}$band,
$f$ is the relative number of holes in the Brillouin zone,
\textit{i.e.} the hole filling factor defined as the ratio of the area
of the hole pocket around the $(\pi,\,\pi)$ point and the area of the
Brillouin zone is $(2\pi)^2$.
For the averaged number of electrons we have
\be
\langle\hat{N}\rangle/{\mathcal N}=2\times2+ 2(1-f).
\ee
For $f=\frac12$ we have a pattern insulator,
while for optimal doping we have
\be
f_\mathrm{opt}=\frac12+0.08.
\ee
The single electron component of the Hamiltonian is diagonalized in
the $\mathbf{p}$-representation and we have to perform the summation
over all four bands $b=1,2,3,4$
\begin{align}
\hat{H}^{\prime (0)}
&=\sum_{\mathrm{b},\mathbf{p},\alpha}(\epsilon_{\mathrm{b},\mathbf{p}}-\mu)
\hat{c}_{\mathrm{b},\mathbf{p},\alpha}^\dagger
\hat{c}_{\mathrm{b},\mathbf{p},\alpha}\notag\\
&\rightarrow
\sum_{\mathbf{p},\alpha}(\epsilon_{\mathbf{p}}-\mu)\,
\hat{c}_{\mathbf{p},\alpha}^\dagger
\hat{c}_{\mathbf{p},\alpha} .
\end{align}
The spectrum $\mathbf{p}$ in the LCAO approximation is determined  by the secular equation
\begin{equation*}
\mathrm{det}[H_\mathrm{LCAO}-\epsilon_\mathbf{p}\openone]
=\mathcal{A}xy+\mathcal{B}(x+y)+\mathcal{C}=0,
\end{equation*}
where the variables 
\begin{equation*}
	x=\sin^2\left(\frac{p_x}{2}\right),\qquad
	y=\sin^2\left(\frac{p_y}{2}\right),
\end{equation*}
are functions only of the momentum $\mathbf{p}$, and
\begin{align*}
\mathcal{A}&=32\left(t_{sp}^2-\frac12\varepsilon_st_{pp}\right)^2
	(2 t_{pd}^2 + t_{pp} \varepsilon_d ), \\
\mathcal{B}&=-4\varepsilon_\mathrm{p}(t_{sp}^2\varepsilon_d+t_{pd}^2\varepsilon_s),\nn \\
\mathcal{C}&=\varepsilon_d\varepsilon_s\varepsilon_\mathrm{p}^2, 
\end{align*}
are polynomials of the energy $\epsilon$.
This analytical representation of the energy dispersion $\epsilon_\mathbf{p}$
allows to express explicitly its derivatives 
\be
\mathbf v_\mathbf{p}=\frac{\partial\epsilon_\mathbf{p}}{\partial\mathbf{p}},\quad
\mathbf{V}_\mathbf{p}=\frac{a_0}{\hbar}\mathbf v_\mathbf{p}=\frac{\md\mathbf{r}}{\md t},\quad
\mathbf{P}=\frac{\hbar}{a_0}\mathbf{p},
\ee
where $\mathbf{P}$ is dimensional momentum and $V_\mathbf{p}$ 
is the velocity in the usual units length per time.

In order to study the low frequency electronic processes,
we restrict the 
Hamiltonian summation only to the conduction band $b=3$ and the index 
will be dropped; $b=1$ and $b=2$ denote completely filled oxygen bands,
while $b=4$ is the index for the completely empty Cu$4s$ band.
Performing this first reduction,
the interaction $s$-$d$-exchange Hamiltonian reads
\be
\hat H_{sd}=-\frac{J_{sd}}{\mathcal{N}} \! \! \!
\sum_{\substack{
\mathbf{p}^\prime+\mathbf{q}^\prime
=\mathbf{p}+\mathbf{q} \\
\alpha, \, \beta}} \! \! \!
S_{\mathbf{q}^\prime} D_{\mathbf{p}^\prime}
{\color{blue}\hat c^\dagger_{\mathbf{q}^\prime\beta} \hat c^\dagger_{\mathbf{p}^\prime\alpha}}
{\color{red}\hat c_{\mathbf{p}\alpha}\hat c_{\mathbf{q}\beta}}
S_\mathbf{p}D_\mathbf{q},
\label{H_sd_1}
\ee
where the real amplitudes
\be
\begin{pmatrix}
\tilde D_\mathbf{p}\\ \tilde S_\mathbf{p}\\ \tilde X_\mathbf{p}\\ \tilde Y_\mathbf{p}
\end{pmatrix}
\! \! =\! \!
\begin{pmatrix}
-\varepsilon_s\varepsilon_\mathrm{p}^2+4\varepsilon_pt_{sp}^2 (x+y)-32t_{pp}\tau_{sp}^2xy\\
-4\varepsilon_\mathrm{p} t_{sp} t_{pd}(x-y)\\
-(\varepsilon_s\varepsilon_\mathrm{p}-8\tau_{sp}y)t_{pd}s_x\\
(\varepsilon_s\varepsilon_\mathrm{p}-8\tau_{sp}x)t_{pd}s_y \nn
\end{pmatrix}
\ee
describe the amplitude of a band electron to be
projected on Cu4$s$, Cu3$d_{x^2-y^2}$, O2$p_x$ and O2$p_y$.
For brevity we introduce the notations 
$\varepsilon_s=\epsilon-\epsilon_s$,
$\varepsilon_d=\epsilon-\epsilon_d$,
$\varepsilon_\mathrm{p}=\epsilon-\epsilon_\mathrm{p}$.
The amplitudes have to be normalized to
$$
C_\Psi=\frac{1}{\sqrt{D_\mathbf{p}^2+S_\mathbf{p}^2+X_\mathbf{p}^2+Y_\mathbf{p}^2}},
$$
and finally  
$S_\mathbf{p} = C_\Psi \tilde S_\mathbf{p}$,
$D_\mathbf{p} = C_\Psi \tilde D_\mathbf{p}$.
For convenience we introduce the notation describing the $s$-$d$ hybridization amplitude $\chi_\mathbf{p}\equiv S_\mathbf{p}D_\mathbf{p}$
of the band electron.
In the next subsection we juxtapose different further reductions
of the exchange Hamiltonian treated in a self-consistent way.

\subsection{BCS versus Fermi liquid reduction}

Our first step is to perform BCS reduction of the exchange Hamiltonian 
\Eqref{H_sd_1}.
In the sum we have to take into account only the
annihilation and creation operators with opposite momenta
and spins
\begin{align}
{\color{blue}\mathbf{q}^\prime = -\mathbf{p}^\prime}, \qquad
{\color{red}\mathbf{q}=-\mathbf{p}},
\qquad \beta=\overline\alpha.
\end{align}
In averaging of this BCS reduced Hamiltonian 
we apply the self-consistent approximation
\begin{align*}
\langle{\color{blue}\hat c^\dagger_{\mathbf{q}^\prime\beta} \hat c^\dagger_{\mathbf{p}^\prime\alpha}}
{\color{red}\hat c_{\mathbf{p}\alpha}\hat c_{\mathbf{q}\beta}}\rangle
	& \rightarrow
\delta_{\mathbf{q}^\prime+\mathbf{p}^\prime,0}\,
\delta_{\mathbf{q}+\mathbf{p},0} \,         
\langle{\color{blue}\hat c^\dagger_{\mathbf{q}^\prime\beta} \hat c^\dagger_{\mathbf{p}^\prime\alpha}}
{\color{red}\hat c_{\mathbf{p}\alpha}\hat c_{\mathbf{q}\beta}}\rangle\nn\\
&
\approx 
\delta_{\mathbf{q}^\prime+\mathbf{p}^\prime,0}\,
\delta_{\mathbf{q}+\mathbf{p},0}\,         
\langle{\color{blue}\hat c^\dagger_{\mathbf{q}^\prime\beta} \hat c^\dagger_{\mathbf{p}^\prime\alpha}}
\rangle\langle
{\color{red}\hat c_{\mathbf{p}\alpha}\hat c_{\mathbf{q}\beta}}\rangle\nn\\
&=\langle{\color{blue}\hat c^\dagger_{-\mathbf{p}^\prime\overline\alpha} 
\hat c^\dagger_{\mathbf{p}^\prime\alpha}}\rangle
\langle{\color{red}\hat c_{\mathbf{p}\alpha}\hat
	c_{-\mathbf{p}\overline\alpha}}\rangle \notag 
=\langle \hat B_\mathbf{p^\prime}\rangle \langle \hat B_\mathbf{p}\rangle,
\end{align*}
where
\be
\hat B_\mathbf{p} \equiv\hat c_{-\mathbf{p},-}\hat c_{\mathbf{p}+}
=u_\mathbf{p}v_\mathbf{p}(1-\hat b_{-\mathbf{p},-}^\dagger\hat b_{-\mathbf{p}-}
-\hat b_{\mathbf{p},+}^\dagger\hat b_{\mathbf{p}+})+\dots \;.\nn
\ee
Colors (on-line) describe factorization of means in the effective Hamiltonian.
The self-consistent approximation is reduced to substitution of averaged product
of four operators to the product of averaged two operators. 
In order to emphasize the basis of the BCS approximation 
we use different colors.
Those colors can be traced back to the conduction band 
reduced exchange Hamiltonian \Eqref{H_sd_1}.
We use standard notations for Bogolyubov $u$-$v$ rotation and 
the new $\hat B$ operators with average expressed in terms of new Fermion number operators $\hat n_{\mathbf{p},\alpha}=\hat b_{\mathbf{p}\alpha}^\dagger\hat b_{\mathbf{p}\alpha}$.
In this way the average exchange Hamiltonian is incorporated in the standard BCS scheme
\begin{equation}
\langle\hat H_{sd}^{\mathrm{(BCS)}}\rangle=
\frac{1}{\mathcal{N}} 
\sum_{\mathbf{p}^\prime,\,\mathbf{p}} 
\langle \hat B_\mathbf{p^\prime}\rangle 
f(\mathbf{p}^\prime,\mathbf{p})
\langle \hat B_\mathbf{p}\rangle,
\end{equation}
where the kernel
\begin{equation*}
f(\mathbf{p}^\prime,\mathbf{p})=-2J_{sd}\chi_{\mathbf{p}^\prime}\chi_{\mathbf{p}}.
\end{equation*}
is separable due to the fact that the exchange interaction is localized on a single ion in the elementary cell of the crystal.

Now we address the Fermi liquid reduction of the same exchange Hamiltonian
\Eqref{H_sd_1} which we rewrite
\be
\hat H_{sd}=-\frac{J_{sd}}{\mathcal N} \!\!
\sum_{\substack{
\mathbf{p}^\prime+\mathbf{q}^\prime=\mathbf{p}+\mathbf{q}\\
\alpha,\, \beta}} \!\!
S_{\mathbf{q}^\prime} D_{\mathbf{p}^\prime}
{\color{red}\hat c^\dagger_{\mathbf{q}^\prime\beta}} 
{\color{blue}\hat c^\dagger_{\mathbf{p}^\prime\alpha}\hat c_{\mathbf{p}\alpha}}
{\color{red}\hat c_{\mathbf{q}\beta}}
S_\mathbf{p}D_\mathbf{q}.
\nn
\ee
In order to point out the difference between BCS and Fermi liquid reduction
now the colors mark the operators which will be grouped in the next self-consistent
approximation.
In the Fermi-liquid (FL) reduced Hamiltonian we have to take into account only the terms with
\be
{\color{blue}\mathbf{p}^\prime = \mathbf{p}}, \qquad
{\color{red}\mathbf{q}^\prime=\mathbf{q}}.
\ee
In FL reduction we have again to apply the self-consistent approximation for the relevant terms
\begin{align*}
\langle{\color{red}\hat c^\dagger_{\mathbf{q}^\prime\beta}} 
{\color{blue}\hat c^\dagger_{\mathbf{p}^\prime\alpha}\hat c_{\mathbf{p}\alpha}}
{\color{red}\hat c_{\mathbf{q}\beta}}\rangle
	&\rightarrow
\delta_{\mathbf{q}^\prime,\mathbf{q}}\,
\delta_{\mathbf{p}^\prime,\mathbf{p}} \,     
\langle{\color{blue}\hat c^\dagger_{\mathbf{p}^\prime\alpha}\hat c_{\mathbf{p}\alpha}}{\color{red}\hat c^\dagger_{\mathbf{q}^\prime\beta}} 
{\color{red}\hat c_{\mathbf{q}\beta}}\rangle\nn\\
&
\approx\delta_{\mathbf{q}^\prime,\mathbf{q}}\,
\delta_{\mathbf{p}^\prime,\mathbf{p}} \,     
\langle{\color{blue}\hat c^\dagger_{\mathbf{p}^\prime\alpha}\hat c_{\mathbf{p}\alpha}}\rangle
\langle{\color{red}\hat c^\dagger_{\mathbf{q}^\prime\beta}} 
{\color{red}\hat c_{\mathbf{q}\beta}}\rangle\nn\\
&
=\langle{\color{blue}\hat c^\dagger_{\mathbf{p}\alpha}\hat c_{\mathbf{p}\alpha}}\rangle
\langle{\color{red}\hat c^\dagger_{\mathbf{q}\beta}} 
{\color{red}\hat c_{\mathbf{q}\beta}}\rangle 
=\langle n_{\mathbf{p},\alpha}\rangle \langle n_{\mathbf{q},\beta}\rangle .\nn
\end{align*}
Analogously to the BCS reduction, now for the FL reduction we
obtain the averaged Hamiltonian
\begin{equation}
	\label{FLH}
\langle\hat H_{_\mathrm{sd}}^{\mathrm{(FL)}}\rangle
=\frac{1}{\mathcal N}\sum_{\mathbf{p},\mathbf{q},\,\alpha,\beta}
\langle \hat n_{\mathbf{p},\alpha} \rangle
f(\mathbf{p},\mathbf{q})
\langle \hat n_{\mathbf{q},\beta} \rangle,
\end{equation}
which is expressed by the same kernel 
\begin{equation*}
f(\mathbf{p},\mathbf{q})=-2J_{sd}\chi_\mathbf{p}\chi_{\mathbf{q}}, \nn\\
\end{equation*}
applied between the average numbers of the Fermi particles
\begin{align*}
\hat n_{\mathbf{p},\alpha} & =\hat c^\dagger_{\mathbf{p\alpha}}\hat c_{\mathbf{p\alpha}},\\
\langle\hat n_{\mathbf{p},\alpha}\hat n_{\mathbf{q},\beta}\rangle &
\approx \langle\hat n_{\mathbf{p},\alpha}\rangle \langle\hat
	n_{\mathbf{q},\beta}\rangle.
\end{align*}

This coincidence of the kernels of BCS and FL approach is one of the central results of the present study.
This coincidence can be explored for application in the study of
other layered transition metal perovskites as well.

\begin{figure}[h!]
\centering
\includegraphics[width=\columnwidth]{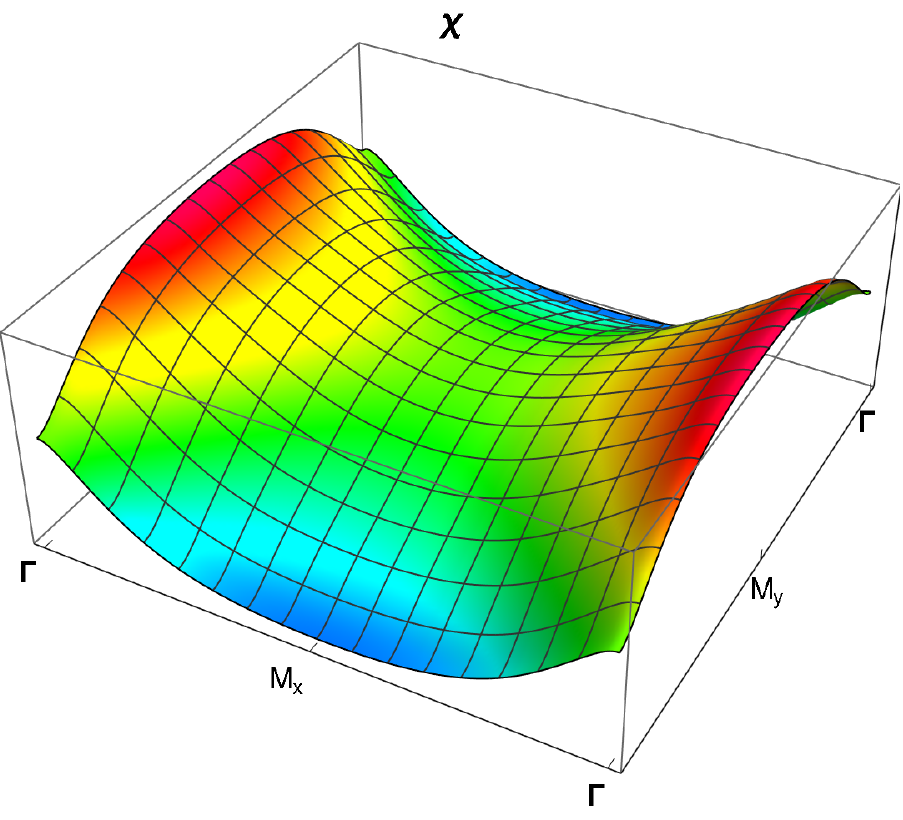}
\caption{The hybridization function $\chi_\mathbf{p}=S_p D_p$
representing the amplitude electron from conduction 
Cu3$d_{x^2-y^2}$ band being simultaneously a Cu$4s$ electron.
This is the main ingredient 
of the matrix elements of the \textit{s-d} exchange interaction.}
\label{fig:chiPlot}
\end{figure}

We wish to emphasize that for the interaction kernel we have
analytical results at hand and for the $s$-$d$ hybridization function,
we have~\cite{MishonovPenev:11}
\begin{align}
\chi_\mathbf{p} = \ & S_\mathbf{p}D_\mathbf{p} \notag \\
= \ & 4\varepsilon_\mathrm{p}t_{sp}t_{pd}(x-y)  \nn \\
& \times \left[
\varepsilon_s\varepsilon_\mathrm{p}^2-4\varepsilon_\mathrm{p}t_{sp}^2\,(x+y)
+32t_{pp}\tau_{sp}^2\,xy
\right]\nn\\
& \times\left\{
\left[4\varepsilon_\mathrm{p}t_{sp}t_{pd}\,(x-y)\right]^2\right.\nn\\
&\qquad
+\left[\varepsilon_s\varepsilon_\mathrm{p}^2-4\varepsilon_\mathrm{p}t_{sp}^2\,(x+y)
+32t_{pp}\tau_{sp}^2\,xy \right]^2\nn\\
&\qquad
+4x\left[(\varepsilon_s\varepsilon_\mathrm{p}-8\tau_{sp}^2y)t_{pd}\right]^2\nn\\
&\qquad
\left.+4y\left[(\varepsilon_s\varepsilon_\mathrm{p}-8\tau_{sp}^2x)t_{pd}\right]^2
\right\}^{-1}.
\label{hybfunc}
\end{align}
This hybridization function in the quasi-momentum representation is depicted in \Fref{fig:chiPlot}.
One can see that at fixed electron energy this saddle can be
approximated by single sinusoidal approximation
$\chi\propto\cos(2\theta)$ where $\theta=\arctan(p_y,p_x)$.
Close to the $(\pi,\,\pi)$-point the single particle spectrum $\epsilon_\mathbf{p}$
has a parabolic form
$\epsilon_\mathbf{p}\approx \epsilon_{\pi,\,\pi}-p^2/2\tilde m_\mathrm{eff}$
and qualitatively this approximation can be extended to the Fermi contour of the 
optimally doped cuprates.

\begin{figure}[ht]
\centering
\includegraphics[width=\columnwidth]{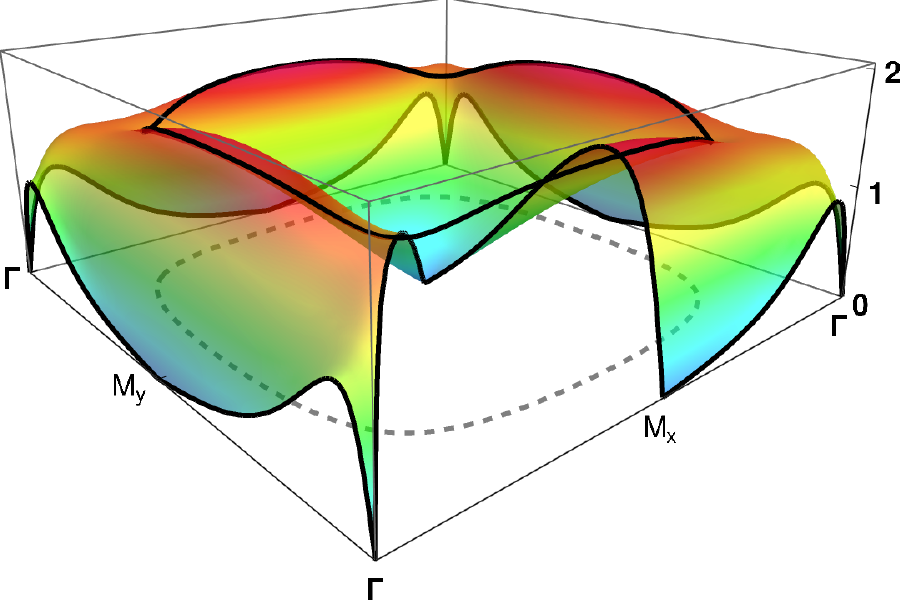}
\caption{Velocity $v_\mathbf{p}$ 
of the conduction band
as a function of quasi-momentum $p_x,\,p_y\in (0,2\pi)$
with dimension energy given in eV.
The variable $V=(a_0/\hbar)v$ has dimension m/s.
In the special points
$\Gamma=(0,\,0)$,
$\mathrm{M}=(\pi,\,0)$,
$\mathrm{X}=(\pi,\,\pi)$
the band velocity $\mathbf V=\partial\epsilon_\mathbf{P}/\partial \mathbf{P}$
is zero; $\mathbf{P}=(\hbar/a_0)\mathbf{p}$.}
\label{fig:cross}
\end{figure}

The band velocity $\mathbf{v}=\partial_\mathbf{p}\epsilon_\mathbf{p}$ 
calculated from the real LCAO Hamiltonian is drawn in \Fref{fig:cross}.
Here we can see that the circular Fermi contour is only a rough initial approximation.

The averaged BCS Hamiltonian has to be minimized with respect to $u_\mathbf{p}$
and then the BCS spectrum 
\be
E_\mathbf{p}
=\frac{\partial\langle \hat H^\mathrm{(BCS)}\rangle}
{\partial {\langle n_{\mathbf{p},\alpha}\rangle}}
=\sqrt{(\epsilon_\mathbf{p}-\mu)^2+\Delta_\mathbf{p}^2}.
\ee

In the next Section we analyze the possible propagation of zero sound
in layered perovskites using the single particle spectrum
obtained from the FL reduced Hamiltonian \Eqref{FLH}.

\section{Zero sound dispersion}

For a concise introduction to the Fermi liquid approach
we recommend the well-known monographs
by Nozieres~\cite{Nozieres}, Abrikosov~\cite{Abrikosov},
Abrikosov, Gor'kov and Dzyaloshinski~\cite{AbrGorDzya},
Lifshitz and Pitaevskii~\cite{LL9} and \cite[Sec.~76]{LL10}.

Introducing for brevity 
$n_{\mathbf{p},\alpha}\equiv\langle\hat n_{\mathbf{p},\alpha}\rangle$,
the averaged FL Hamiltonian \Eqref{FLH} reads
\be
H_\mathrm{_{FL}} \equiv
\langle \hat H^{\mathrm{(FL)}}\rangle 
= \sum_{\mathbf{p},\,\alpha}\epsilon_\mathrm{p}n_{\mathrm{p},\alpha}
+\frac{1}{\mathcal N}\!\sum_{\mathbf{p},\mathbf{q},\,\alpha,\beta}
n_{\mathbf{p},\alpha} f(\mathbf{p},\mathbf{q})
n_{\mathbf{q},\beta}.
\ee
Notice that the spin indices may be omitted if we consider spin
non-polarized phenomena,
such that $n_{\mathbf{q},+} = n_{\mathbf{q},-}$. Introducing
$n_{\mathbf{q}} \equiv n_{\mathbf{q},+}+n_{\mathbf{q},-} =
2 n_{\mathbf{q},+} =  2 n_{\mathbf{q},-}$
for the FL energy spectrum we get
\begin{align}
\varepsilon(\mathbf{p},\mathbf{r})
= \frac{\partial H_{_\mathrm{FL}}}{\partial\hat n_{\mathbf{p}}} 
	= \ & \epsilon_\mathbf{p}+\frac1{\mathcal N}\sum_{\mathbf{q}}f(\mathbf{p},\mathbf{q}) 
n_{\mathbf{q}}(\mathbf{r})\nn\\
= \ &\epsilon_\mathbf{p}
+\frac{(-2J_{sd})}{\mathcal N}\chi_\mathbf{p}\sum_\mathbf{q}\chi_\mathbf{q} 
n_{\mathbf{q}}(\mathbf{r},t),
\label{Fermi_liquid_spectrum}
\end{align}
where the space variable $\mathbf{r}\equiv a_0\mathbf{n}$
can be introduced only in the quasi-classical WKB approximation.
The $s$-$d$ hybridization function \Eqref{hybfunc} determines both the
gap anisotropy $\Delta_\mathbf{p}(T)=\Xi(T)\chi_\mathbf{p}$ and the FL interaction.
A detailed analysis of the gap anisotropy and 
the hot/cold spot anisotropy of ARPES data
described by the hybridization function $\chi_\mathbf{p}$
in \Eqref{hybfunc} is presented in Ref.~\cite{hotspot2022}.
Now we consider the momentum distribution $n_{\mathbf{p}}(\mathbf{r},t)$ 
of the charge carriers as dynamic variables. 
Assuming a local deformation of the Fermi contour in two dimensions
\begin{align*}
	n_{\mathbf{p}}(\mathbf{r},t) & =2\,
\theta(\eF+\nu_\mathbf{p}(\mathbf{r},t)-\epsilon_\mathbf{p}),\\
n_{\mathbf{p},+}^{(0)} & =n_{\mathbf{p},-}^{(0)}
=\theta(\eF-\epsilon_\mathbf{p}),\nn\\
 \delta n_{\mathbf{p}}
& =n_{\mathbf{p}}(\mathbf{r},t)-n_{\mathbf{p}}^{(0)}\nn
\end{align*}
and using the collisionless Boltzmann equation,
we derive the integral equation for the deformation of the Fermi contour.
In the linearized with respect to small $\nu$ equation we assume
plane wave perturbations, \textit{i.e.}
\begin{equation}
\label{deformation}
 \delta n_{\mathbf{p}}(\mathbf{r},t)\approx 2\,
\delta(\eF-\epsilon_\mathbf{p})
\nu_\mathbf{p} 
\exp[\mathrm{i} (\mathbf{K}\cdot\mathbf{r}-\omega t)],
\end{equation}
with wavevector $\mathbf{K}=\mathbf{k}/a_0$
and frequency $\omega$.

The WKB energy $\varepsilon(\mathbf{p},\mathbf{r})$ from \Eqref{Fermi_liquid_spectrum}
is actually an effective Hamiltonian which gives the force 
acting on the electrons
\be
\mathbf{F}_{\! \mathbf{p}}(\mathbf{r},t)
=-\frac{\partial \varepsilon(\mathbf{p},\mathbf{r})}{\partial \mathbf{r}}
=\frac{\md\mathbf{P}}{\md t}
\ee
and together with the substitution \Eqref{deformation} of $n_{\mathbf{p}}(\mathbf{r},t)$
gives
\be
\mathbf{F}_{\! \mathbf{p}}(\mathbf{r},t)
=-\mathrm{i} \mathbf{K}
\int\limits_{\mathrm{BZ}}\!\! f(\mathbf{p},\mathbf{p}^\prime)\,\delta n_{\mathbf{p}^\prime}
\frac{\md p_x^\prime\md p_y^\prime}{(2\pi)^2}.
\ee
The substitution of the small deformation of the Fermi contour $\delta n_{\mathbf{p}}$  
from \Eqref{Fermi_liquid_spectrum} and \Eqref{deformation} in the collisionless Boltzmann equation
\be
\frac{\md}{\md t} \delta n_{\mathbf{p}}(\mathbf{r},t)
=\!\frac{\partial \delta n_{\mathbf{p}}}{\partial t}
+\frac{\partial \delta n_{\mathbf{p}}}{\partial \mathbf{P}}\cdot \,
\mathbf{F}_{\! \mathbf{p}}
+\frac{\partial \delta n_{\mathbf{p}}}{\partial \mathbf{r}}\cdot 
\mathbf{V}_{\! \mathbf{p}}=0
\ee
after some algebra leads to the dispersion equation
\be
\left(\omega-\mathbf{K}\cdot\mathbf{V}_\mathrm{F}(\mathbf{p})\right)\nu_\mathbf{\mathbf{p}}
=\frac{\mathbf{K}\cdot\mathbf{V}_\mathrm{F}(\mathbf{p})}{(2\pi)^2}
\oint\limits_\mathrm{FC}  f(\mathbf{p},\mathbf{p}^\prime)
\nu_{\mathbf{p}^\prime}\frac{\md p_l^\prime}{v_\mathrm{_F}(p_l^\prime)},
\ee
where
\begin{align}
K_x=K\cos{\beta},\quad K_y=K\sin{\beta},\quad\mathbf{k}=a_0 \mathbf{K}.\nn
\end{align}

To proceed further we introduce the averaging on the Fermi contour
\begin{equation*}
\langle F \rangle \equiv \frac1{\rho_{_F}}\oint\limits_\mathrm{FC}
F(\mathbf{p}) \, \frac{\md p_l}{(2\pi)^2 v_\mathrm{_F} (\mathbf{p})},\\
\end{equation*}
where 
\begin{equation*}
\rho_{_F}\equiv \oint\limits_\mathrm{FC}
1 \, \frac{\md p_l}{(2\pi)^2v_\mathrm{_F} (\mathbf{p})}
\end{equation*}
 is the density of electronic states  per elementary cell and spin.
For the separable kernel of the $s$-$d$ interaction
the dispersion equation for the zero sound reads
\begin{align}
\hbar\omega\langle\chi\nu\rangle-\mathbf{k}\cdot\langle\mathbf{v}\chi\nu\rangle
=(-2J_{sd})\rho_{_F}\mathbf{k}\cdot\langle\mathbf{v}\chi^2\rangle
\langle\chi\nu\rangle,
\nn
\end{align}
where the momentum argument of all those average values is omitted for brevity.
In order to analyze qualitatively the solutions of this integral equation for $\nu(\mathbf{p})$ 
we approximate the Fermi contour with a circle
$\epsilon_\mathbf{p}=\mathbf{P}^2/2m_\mathrm{eff}$ and
assume that the separable kernel behaves as a single sinusoidal on the Fermi contour
\begin{equation}
\label{sinus}
V_{\mathbf{p},\mathbf{q}^\prime}\approx I_{sd}\,\cos(2\theta)\cos(2\theta^\prime),
\end{equation}
where
\begin{align*}
& I_{sd}\approx -2J_{sd}\,\left[\frac{t_{sp}t_{pd}}
{(\epsilon_s-\epsilon_d)(\epsilon_d-\epsilon_\mathrm{p})}\right]^2, \\
& p_x=p\cos{\theta},\quad p_y=p\sin{\theta},\\
& q_x^\prime=q^\prime\cos{\theta^\prime},\quad q_y^\prime=q^\prime \sin{\theta^\prime},\\
& \frac{p^2}{2\tilde{m}_\mathrm{eff}}=\frac{q^2}{2\tilde{m}_\mathrm{eff}}=\eF.
\end{align*}
It is worth mentioning that many authors just postulate such a separable
kernel for the pairing interaction $\chi\propto\cos(2\theta)$,
while we have derived it from the $s$-$d$ microscopic Hamiltonian.

If we analyze the propagation of zero sound along the nodal lines
of the hybridization function $\beta=\pi/4$ the electric current 
and charge density oscillations are zero.
In the general case for significant charge density oscillations 
the zero sound is actually a plasmon.

\begin{figure}[t!]
\centering
\includegraphics[scale=0.5]{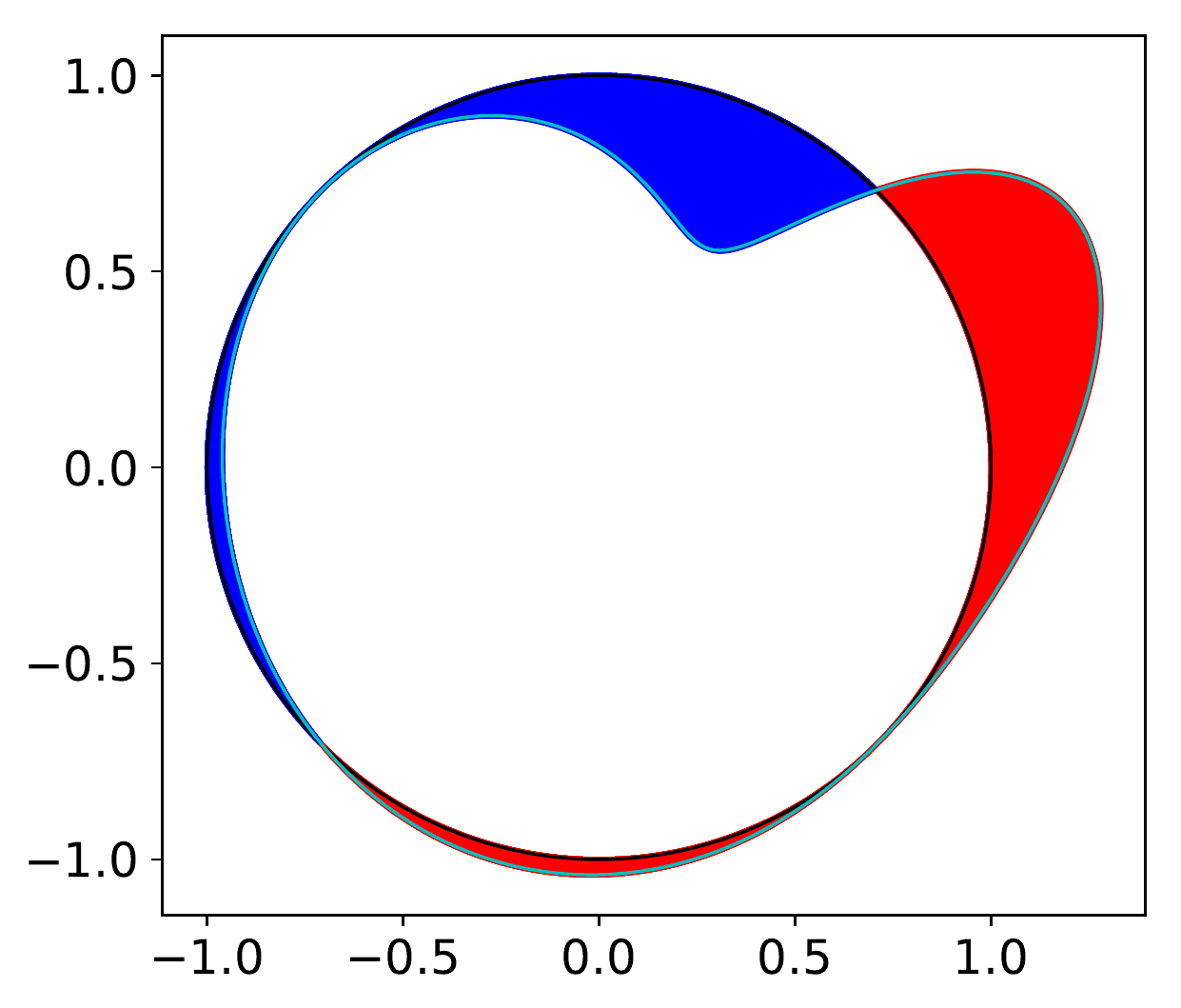}
\caption{Deformation of the Fermi contour in two dimensional 
momentum space $\mathbf{p}/\pi$ for zero sound 
propagating along
``cold spots''~\cite{Ioffe:98} diagonal $\beta=\pi/4$ 
in layered perovskites.
For this special case of propagation along the nodal
lines of the separable kernel of interaction,
the electric charge, spin and current oscillations are zero.
Moreover,
we have a shear deformation of the Fermi surface in momentum space.
}
\label{fig:pimpo}
\end{figure}

The deformation of the Fermi circle in this special case $\beta=\pi/4$
is depicted in \Fref{fig:pimpo}.
For propagation of zero sound along other directions
it is necessary to take into account the Coulomb repulsion 
and zero sound will behave partially as a plasmon in a layered structure.
For a general review on plasmons in cuprate superconductors see
Ref.~\cite{Greco:19}.

\section{Conclusion and discussion}

We investigated the propagation of the zero sound in a class of layered
transition metal perovskites involving a $s$-$d$ interaction.
We started and focused our attention on the CuO$_2$ plane
that is a well-studied system. We were able to compare BCS and
Fermi liquid reductions of the 
Hamiltonian and as a property of CuO$_2$ plane these two model Hamiltonian coincided.
Moreover, we studied the influence of the sign of $J_{sd}$ coupling. 
For anti-ferromagnetic sign we have tendency to superconductivity,
while for ferromagnetic sign we expect zero sound to be observed.

Unfortunately at the moment,
starting with a theoretical scheme, it is difficult to conclude in which 
layered compounds the zero sound will have the longest lasting propagation and
what material is technologically suitable to produce a clean
$ac$-surface. It is most likely that a thin layer geometry would
provide a solution.


Owing to the research outlined throughout this paper we conclude:
First of all, zero sound exists only 
when $J_{sd}$ is negative and the $s$-$d$ interaction has a ferromagnetic sign.
However, the $s$-$d$ exchange can create superconductivity only for antiferromagnetic  
sign (positive $J_{sd}$) of the exchange interaction.
We arrive to the conclusion that 
in the normal phase of high-$T_c$ cuprates propagation of zero sound is impossible.
Zero sound for high-$T_c$ cuprates is a dissipation mode, 
but thermal excitation of all those modes creates 
intensive scattering and Ohmic resistivity due to the exchange interaction
and this strong angular dependence of the scattering rate is the cause of so called 
``hot spots'' phenomenologically postulated for the interpretation of the experimental data~\cite{Hlubina:95}.
Here we wish to add that thermal fluctuations of plasmons could also contribute to the
hot spots along the Fermi contour
\cite{Greco:19}.

Thermally agitated plasmons are related to electron density fluctuations which create electron scattering and ohmic resistivity due to exchange interaction.
However, not for all doping levels the cuprates are superconducting and we do not exclude 
 $J_{sd}$ to change its sign for some compounds.

Our main motivation to write this paper is to attract the attention
of experimentalists 
with appropriate samples at hand to probe the zero-sound propagation in the $ab$-plane of
transition metal layered perovskites. 
If Angle Resolved Photoemission Spectroscopy (ARPES) data are available for these materials,
hot spots along the Fermi contour or even smearing of this contour
will be a significant hint for intensive $s$-$d$ exchange 
which can lead to propagation of zero sound.
In normal metals,
the anisotropy of the electron-electron interaction is not strong enough
to ensure zero sound propagation,
but for layered perovskites such a phenomenon is most likely to occur.
Another hint for intensive $s$-$d$ exchange can come from band calculations,
the hybridization is strongest if all those 3 levels: for transition ion
4$s$ and 3$d$ and $p$-states for the chalcogen are close to each other and we have 
almost a triple coincidence (full overlapping).

From the practical point of view, a possible route towards the excitation of the zero sound
could be achieved by an intense perturbation from one side of a narrow strip,
and detection on the opposite side of the sample. This, however is
still a remote possibility.

Last but not least, already a half century ago
different kinds of zero sound are extensively studied by theoretical
means. This topic continues to attract a great deal of interest within the
scientific community. Here we mention but a few papers that are
somehow directly linked to our study.
Recent considerations include the two-dimensional zero-sound~\cite{Khoo:19}
and shear~\cite{Marel:21} zero sound for $p$-type interaction~\cite{Ding:19},
and we finally conclude that except for $^3$He
thin films and even two dimensional structures 
with large exchange interaction with ferromagnetic sign soon will become an interesting 
object for realization of the old idea of Landau~\cite{Landau_theory,Landau_zero_sound}.


In this paper we have devised the theoretical framework for the
possible emergence of zero sound in some layered perovskites
involving ferromagnetic $s$-$d$ exchange interaction. We will continue
our effort to extend the investigation to other transition-metal compounds along with
distinct geometries to put the test the plausibility of the present
theory. From the experimental side we hope that the current technological
progress would make it possible to synthesize appropriate compounds
allowing for the propagation of zero sound.

\section*{Acknowledgments}

The authors  are thankful to Davide~Valentinis for the interest to the present study
and pointing out for recently appeared related works on kinetic theories for the 
electrodynamic response of Fermi liquids and anisotropic metals 
\cite{Valentinis:21,Valentinis:22, Baker:22}.

The authors AMV, TMM and NIZ  are grateful to Cost Action CA16218
Nanoscale coherent hybrid devices for superconducting quantum technologies
for the support in presenting the preliminary results of the current research at the
7$^\mathrm{th}$ International Conference on Superconductivity and Magnetism
in Bodrum, Turkey in 2021 and for the interest to the presenting of the final version
at the CA16218 meeting in Madrid, Spain in 2022.
This work is partially supported by grant No K$\Pi$-06-H38/6
of the Bulgarian National Science Fund.



\bibliography{../Article/Pokrovsky}

\end{document}